# Transport Spectroscopy of Sublattice-Resolved Resonant Scattering in Hydrogen-Doped Bilayer Graphene

Jyoti Katoch[1,3]*, Tiancong Zhu[1]*, Denis Kochan[2], Simranjeet Singh[1,3], Jaroslav Fabian[2], and Roland K. Kawakami[1]

[1]*Department of Physics, The Ohio State University, Columbus, Ohio 43210, USA*
[2]*Institute for Theoretical Physics, University of Regensburg, 93040 Regensburg, Germany*
[3]*Department of Physics, Carnegie Mellon University, Pittsburgh, Pennsylvania 15213, USA*

We report the experimental observation of sublattice-resolved resonant scattering in bilayer graphene by performing simultaneous cryogenic atomic hydrogen doping and electron transport measurements in ultrahigh vacuum. This allows us to monitor the hydrogen adsorption on the different sublattices of bilayer graphene without atomic-scale microscopy. Specifically, we detect two distinct resonant scattering peaks in the gate-dependent resistance, which evolve as a function of atomic hydrogen dosage. Theoretical calculations show that one of the peaks originates from resonant scattering by hydrogen adatoms on the α-sublattice (dimer site) while the other originates from hydrogen adatoms on the β-sublattice (non-dimer site), thereby enabling a method for characterizing the relative sublattice occupancy via transport measurements. Utilizing this new capability, we investigate the adsorption and thermal desorption of hydrogen adatoms via controlled annealing and conclude that hydrogen adsorption on the β-sublattice is energetically favored. Through site-selective desorption from the α-sublattice, we realize hydrogen doping with adatoms primarily on a single sublattice, which is highly desired for generating ferromagnetism.

*These authors contributed equally.

Two-dimensional materials are atomically thin membranes with extreme surface sensitivity, which enables unprecedented tuning of electronic, magnetic, and spintronic properties via surface modification [1,2]. In particular, hydrogenation has emerged as a powerful technique to alter the electronic properties and add much sought after magnetism in graphene [3,4]. It has been experimentally observed that hydrogen atom adsorption on single layer graphene can induce magnetic moments [5], increase spin-orbit coupling [6,7], and open a band gap in otherwise gapless graphene [8]. Significantly, these studies have identified the important role of resonant scattering in hydrogenated graphene [3,9-11]. The covalent bonding of hydrogen atoms on graphene produces a localized defect state whose energy lies very close to the Dirac point (or equivalently, the charge neutrality point, CNP). When the Fermi level is tuned into resonance with the defect level, conduction electrons are captured by the localized state to produce strong momentum scattering. Theoretical studies show that this capture process can also affect spin transport in graphene by greatly enhancing spin-relaxation via the exchange coupling with localized magnetic moments and the spin-orbit coupling induced by local curvature [11-13].

However, the direct experimental study of resonant scattering in graphene is very challenging. This is because the defect levels are very close to the CNP, which makes it difficult to resolve the separate gate dependent resistance peaks from the CNP and from resonant scattering. To our knowledge, only one transport study has reported the observation of resonance peaks due to resonant impurities in single layer graphene [14]. However, the use of hydrogen plasma deposition in the study lacks precise control over hydrogenation and may also produce lattice vacancies, which complicates the interpretation of the results.

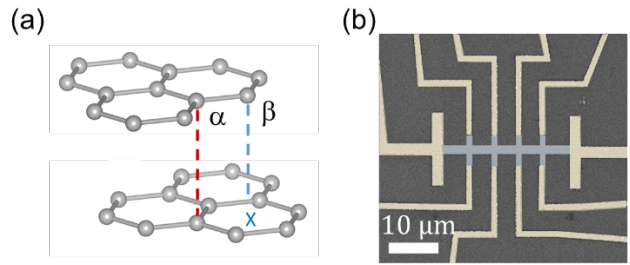

FIG. 1. (a) Schematic of bilayer graphene, depicting α and β sites. (b) SEM picture of the typical device used for in situ hydrogenation and transport measurements in this study.

Even more fascinating, yet still not fully explored is the adatom engineering of the properties of bilayer graphene. Bernal stacked bilayer graphene is a zero gap semiconductor, but breaking the inversion symmetry between the two layers by adatoms [15] or electrostatic gates [16,17] can open up an electronic band gap. Moreover, due to the non-vanishing density of states in bilayer graphene, the electron-hole puddles are less effective in broadening of the Fermi level and therefore the resonant scattering is more pronounced in comparison to single layer graphene. Recently, Kochan *et al.* calculated resonant scattering for hydrogen adatoms on bilayer graphene and predicted two distinct resonances associated with the two inequivalent sublattices: the α-sublattice consisting of carbon atoms in the top layer above filled sites in the bottom layer ("dimer site"), and the β-sublattice consisting of carbon atoms in the top layer above vacant sites in the bottom layer ("non-dimer site") as depicted in Figure 1(a) [18]. Kochan's results suggest that the resonance peaks could be detected in measurements of graphene resistance vs. gate voltage, but such an effect was not observed in experiments [19]. Experimental demonstration of the two resonance peaks would open the exciting possibility of characterizing sublattice occupancy for developing methods [20,21] to generate occupancy of a single sublattice, which is a requirement for realizing ferromagnetic order of the induced magnetic moments [4,20-23].

In this Letter, we perform hydrogen adatom doping of bilayer graphene in ultrahigh vacuum (UHV) and report the observation of sublattice-dependent resonant scattering, as predicted theoretically, through systematic *in situ* transport measurements. By alternating between hydrogen adatom deposition and transport measurements, both performed in UHV at 21 K, we observe the gradual emergence of two additional peaks in the gate-dependent resistance away from the CNP. Using the input from density functional theory (DFT) [18] and Boltzmann transport theory, we are able to identify one of the peaks with resonant scattering induced by hydrogen adatoms on the α-sublattice (dimer site), and the other peak from resonant scattering by hydrogen adatoms on the β-sublattice (non-dimer site). Interestingly, we find that at low dosage atomic hydrogen adsorption onto graphene at 21 K results in a higher occupancy on the β-sublattice compared to the α-sublattice, based on the relative peak heights. Subsequent studies where the sample is heated to an annealing temperature and re-cooled to 21 K for transport measurements reveal characteristics of the thermally-induced diffusion and desorption of the hydrogen adatoms. Between annealing temperature of 21 K and 100 K, the α-peak height decreases gradually while the β-peak remains nearly unchanged. The α-peak virtually disappears when annealed up to 140 K. This suggests that a vast majority of the hydrogen occupies a single sublattice, which is necessary for generating ferromagnetism in hydrogen-doped graphene [4,20-23]. Further increases in annealing temperature produces shifts in the CNP as well as desorption from the β-sublattice, returning the graphene to a nearly pristine (undoped) state. Obtaining such insights on the atomic scale structure from macroscopic transport measurements is quite remarkable and is uniquely enabled by the extreme surface sensitivity of 2D materials.

Bilayer graphene flakes are obtained by mechanically exfoliating Kish graphite onto $SiO_2$(300 nm)/Si substrates. The bilayer thickness is confirmed by Raman spectroscopy [24]. The graphene flakes are patterned into Hall bars using electron beam lithography and reactive ion etch, followed by deposition of Cr (10 nm) /Au (70 nm) contact electrodes. Figure 1(b) shows an SEM image of a device with channel length and width of 5 μm and 2 μm respectively. To remove lithography resist residue, the devices are annealed in $Ar/H_2$ atmosphere for 1 hour at 300°C [25]. Subsequently, the bilayer graphene device is transferred in an ultra-high vacuum (UHV) chamber with base pressure $< 1 \times 10^{-10}$ torr for in situ transport measurements. The devices are annealed at 150°C for 3 hours to remove ambient adsorbates and then cooled down to 21 K for transport measurements. The four-probe resistance, R is measured as a function of back gate voltage, $V_g$, for pristine (undoped) bilayer graphene prior to hydrogenation. The peak in the *R vs. $V_g$* curve occurs when the Fermi level coincides with the CNP of bilayer graphene, and the corresponding gate voltage is denoted as $V_{CNP}$.

We measured the impact of atomic hydrogen adsorption and dehydrogenation on four bilayer graphene devices, and they all showed consistent behavior. We present data from a representative bilayer graphene device. The initial calculated electron and hole field effect mobilities of 3380 $cm^2$/Vs and 2530 $cm^2$/Vs, respectively, are based on the slope of *1/R vs. $V_g$* away from the CNP. To perform hydrogenation, the bilayer graphene device is exposed to atomic hydrogen flux at 21 K (Omicron atomic hydrogen source at 60 W, 3" distance, and $H_2$ background pressure of $1 \times 10^{-7}$ torr) [5]. We alternate between hydrogen dosing and in situ measurement of *R vs. $V_g$* to track the evolution of transport properties with increasing hydrogen adatom coverage. Figure 2(a) shows the gate dependent resistance of the bilayer graphene device for a series of total hydrogenation times. For very low atomic hydrogen dosage (1 s as shown in Figure 2(a)), the graphene resistance increases and $V_{CNP}$ shifts slightly toward negative gate voltages. This behavior is expected, as the hydrogen adatoms will act as scattering centers and donate electrons to the graphene sheet. Similar behavior has also been observed in hydrogenation of single layer graphene [26]. However, after 8 s of total hydrogenation, the behavior of $R$ vs. $V_g$ starts to deviate from the case of single layer graphene. In addition to the increase of resistance and negative shift of the CNP, an extra resistance peak appears on the electron-doping side ($V_g > V_{CNP}$) of the gate dependent resistance curve. With further hydrogen dosage beyond 24 s, we observe the emergence of a second resistance peak appearing between $V_{CNP}$ and the first resistance peak. The two peaks become more prominent with further hydrogenation. As we stop the hydrogenation at 96 s, we clearly see the two resistance peaks, one sharp and one broad, in the $R$ vs. $V_g$ curve away from the CNP peak. In addition, the resistance of the CNP peak increases by an order of magnitude compared to the pristine device. We roughly estimate the atomic hydrogen coverage to be ~0.1% by fitting the conductivity vs. $V_g$ away from the CNP [11,19] (see Supplemental Material for details [27-33]). To better illustrate the evolution of the two resistance peaks, we plot the same data set of gate dependent resistance curves as a function of $V_g - V_{CNP}$ to display the peak positions relative to the CNP (Figure 2(b)). Clearly, with increasing hydrogen dosage, the broader peak labeled "β" emerges first and the sharper peak labeled "α" emerges second. With increasing atomic hydrogen dosage, both peaks shift away from the CNP. The CNP peak, which we label as "CNP" increases in amplitude with increasing hydrogen dosage.

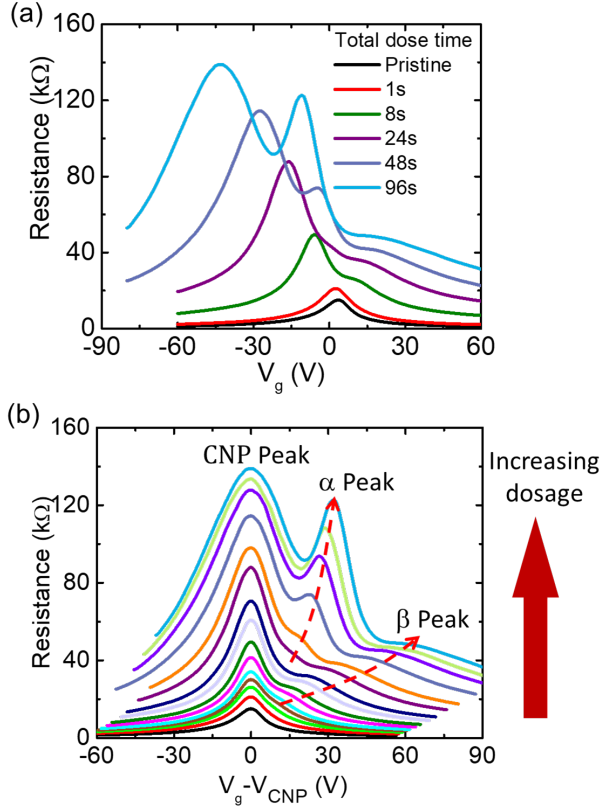

FIG. 2. (a) Gate dependent resistance of undoped (black color curve) bilayer graphene and after exposure to 1 s, 8 s, 24 s, 48 s, and 96 s of total atomic hydrogen dosing times at 21 K. (b) Same experimental dataset in (a), with the evolution of bilayer graphene resistance as a function of $V_g-V_{CNP}$. The bottom curve is for pristine graphene, and the top curve is for 96 s of hydrogen dosage. The dashed arrow lines show the progression of the two peaks labeled as α peak and β peak in resistance with increasing hydrogen dosage.

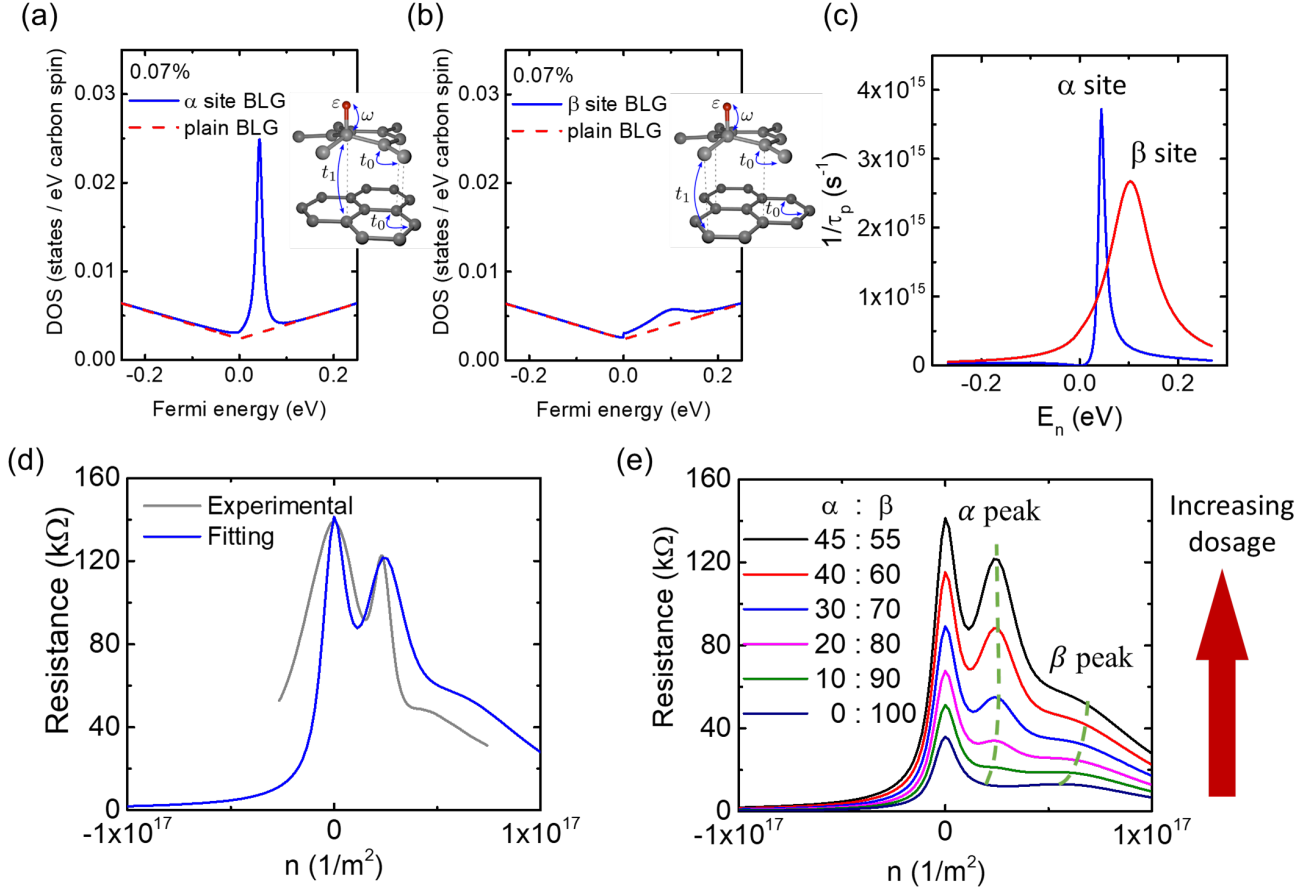

FIG. 3. (a, b, c) Tight-binding model calculation of the density of states (DOS), respectively, for 0.07% hydrogen doping of the α-sublattice (dimer site) and β-sublattice (non-dimer site) and corresponding momentum relaxation-rates. Insets in (a) and (b) show tight-binding parameterization: ε is the on-site energy of the hydrogen orbital, ω is the hybridization between hydrogen $s$ and carbon $p_z$ orbitals, and $t_0$ and $t_1$ are the intralayer and interlayer nearest neighbor hoppings, respectively, in graphene bilayer. (d) Model computed data versus experiment for α:β ratio 45:55, where grey curve is the experimental data in Fig. 2(a) with total hydrogenation of 96 s. (e) Calculation of the bilayer graphene resistance vs. carrier density ($n$ is positive for electrons) for different relative occupancies of the α-sublattice and β-sublattice for α:β ratios ranging from 45:55 to 0:100 to simulate the hydrogenation process. The bottom curves are of lower total hydrogen concentration and the top curves are of higher hydrogen concentration.

As implied by the peak labeling, we attribute the two additional resistance peaks to resonant scattering induced by hydrogen adatoms on different sublattices of the bilayer graphene. Hydrogen adatoms are known to induce resonant impurity states on graphene. In the case of single layer graphene, the energy levels of the defect states induced by hydrogen adatoms are at the Dirac point. In transport measurements, this gives rise to an anomalous large increase in total resistivity at the CNP. For bilayer graphene, due to Bernal stacking, the two sublattices of the top layer are no longer equivalent. This leads to two different resonant impurity states when atomic hydrogen bonds to the top layer. Figures 3(a) and 3(b) show the calculated density of states (DOS) spectra for the resonant defect levels for hydrogen on the β-sublattice and α-sublattice, respectively. The energy level of the defect state on the α-sublattice is closer to the charge neutrality point and has a sharper peak, while the energy level of the defect state on the β-sublattice is farther from the CNP and has a broader peak, both features clearly manifest in momentum relaxation-rate displayed in Figure 3(c). Related zero energy modes in the case of vacancy [34] show the same sharp-broad peak characteristics. This appears to be consistent with characteristics of the two additional peaks in Figure 2(b).

To compare with the gate-dependent resistance data, we apply Boltzmann transport theory based on the tight-binding model [18] and calculate the $R$ vs. $V_g$ peaks for resonant scattering induced by hydrogen. For these calculations [35], we assume the concentration of hydrogen atoms to be 0.07% and vary the relative sublattice occupancy ratios (α:β) to simulate the experimental data. For example, the measured resistance (grey curve) at highest dose time of 96 s is in reasonable agreement with the calculated resistance (blue curve) for relative hydrogen adatom occupancy ratio of 45:55, as shown in Figure 3(d) [36]. To understand the experimental trend observed in Figure 2(b), we plot in Figure 3(e) a set of theoretical resistance curves beginning with α:β of 0:100 for low hydrogen concentration and progressing to 45:55 for high hydrogen concentration (the effect of hydrogen concentration is modeled as a scaling factor for each resistance curve). The evolution of the α and β resonant scattering peaks with increasing hydrogen concentration is similar to the behavior observed in the experimental data (Figure 2(b)). The comparison clearly reveals that there is a much higher occupancy of the β-sublattice as compared to the α-sublattice at low hydrogen dosage, see data in Figure 2(b) between 1 and 24 s. This is expected in the view of previous theoretical calculations predicting that at lower hydrogen concentrations, hydrogen adsorption on the β-sublattice is more stable, with lower energy minima as compared to the α-sublattice [20]. The evolution of the α:β ratio with hydrogen coverage is discussed further in the Supplemental Material [27]).

We now employ the sublattice-resolved transport spectroscopy to investigate the thermally-induced diffusion and desorption of hydrogen on bilayer graphene. This issue is crucial for magnetic ordering of localized moments in graphene, which is predicted to become ferromagnetic if the localized moments lie on the same sublattice [4,20-23]. In fact, sublattice-selective desorption was proposed as a method for achieving hydrogen occupation on a single sublattice [21]. To understand the thermally-induced dynamics, we perform an annealing study by cycling the hydrogenated bilayer graphene device to different elevated temperatures and monitoring the change with transport measurements after cooling back down to 21 K. Fig. 4(a) depicts the schematic for the heating cycle and measurement process. For annealing temperatures increasing from 21 K to 100 K (Fig. 4(b)), there is a strong reduction of the α-peak, which eventually completely disappears at 140 K (Fig. 4(c)). Meanwhile, the broader β-peak and the CNP peak remain largely unchanged. This striking result suggests selective removal of hydrogen adatoms from the α-site and leaving a vast majority of the hydrogen adatoms on a single sublattice (β-site), which is a necessary condition for ferromagnetism. This also supports our earlier conclusion that the β-sublattice has lower energy for hydrogen adsorption compared to the α-sublattice. A detailed discussion about the selective removal of hydrogen adatoms can be found in the Supplemental Material [27].

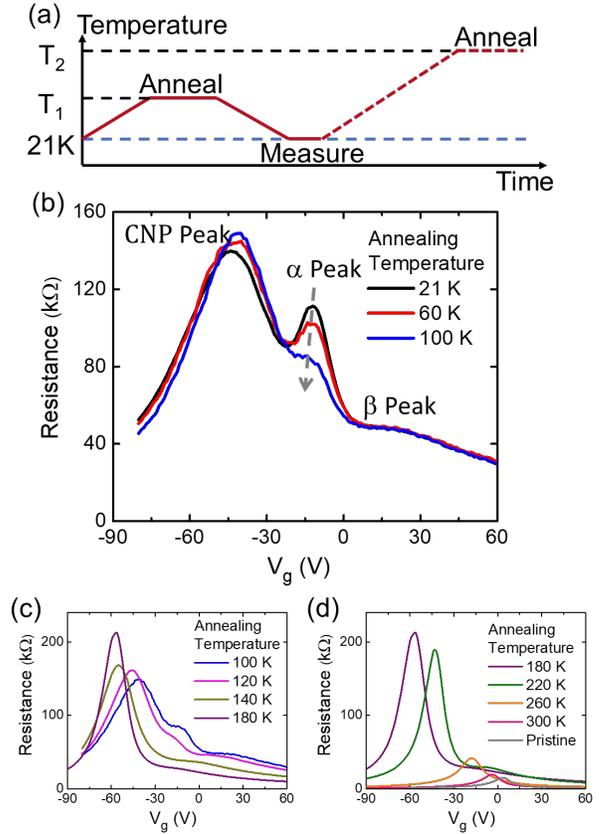

FIG. 4. (a) Diagram of the annealing process for the dehydrogenation study. (b-d) Gate-dependent resistance for annealing temperature from 21 K to 100 K (b), from 100 K to 180 K (c), and from 180 K to 300 K (d). All measurements are taken at 21 K.

Between 100 K and 180 K annealing temperature (Fig. 4(c)), the CNP peak increases substantially and shifts to more negative gate voltages, while the β-peak slightly decreases. This could be due to hydrogen adatoms forming more complex structures, but more detailed experiments using scanning tunneling microscopy would be needed to resolve this issue. Between 180 K and 300 K annealing temperature (Fig. 4(d)), the β-peak disappears and the magnitude of CNP peak decreases and shifts back toward zero gate voltage, which signifies desorption of the hydrogen adatoms. By 300 K annealing temperature, the gate-dependent resistance nearly recovers to a pristine graphene state, indicating the hydrogenation process is reversible. This study illustrates the value of sublattice-resolved transport spectroscopy, which is able to monitor the thermally-induced hydrogen adatom dynamics and map out the characteristic temperatures when various processes are activated.

In conclusion, we utilize *in situ* transport measurements to systematically investigate resonant scattering by hydrogen adatoms on bilayer graphene. These studies realize well-separated peaks in the gate dependent resistance arising from resonant scattering, which has been difficult to achieve experimentally. The observed peaks are attributed to sublattice-dependent resonances as predicted theoretically, and analysis of the resistance curves show that at low dosages the hydrogen adsorbs preferentially to the β-sublattice over the α-sublattice. Using this new capability for sublattice-resolved transport spectroscopy, we investigate the thermally-induced diffusion and desorption of the hydrogen adatoms. Specifically, we find that the thermal desorption of the α-sublattice occurs before the β-sublattice, leading to hydrogen adatoms primarily occupying a single sublattice at intermediate temperatures. This sets the stage for the pursuit of ferromagnetic ordering in hydrogen-doped graphene [4,20-23].

### ACKNOWLEDGEMENTS

J. K. and T. Z. contributed equally to this work. This project was primarily supported by the US Department of Energy (Grant No. DE-SC0018172). J. K. and S. S. acknowledge partial support from the Center for Emergent Materials: an NSF MRSEC under Grant No. DMR-1420451. J. F. and D. K. acknowledge support from the European Union's Horizon 2020 research and innovation program under Grant Agreement No. 696656 and from the DFG SFB 1277 (A09 and B07).

# Supplemental Material for "Transport Spectroscopy of Sublattice-Resolved Resonant Scattering in Hydrogen-Doped Bilayer Graphene"

Jyoti Katoch[1,3]*, Tiancong Zhu[1]*, Denis Kochan[2], Simranjeet Singh[1,3], Jaroslav Fabian[2], and Roland K. Kawakami[1]

[1]*Department of Physics, The Ohio State University, Columbus, Ohio 43210 USA*
[2]*Institute for Theoretical Physics, University of Regensburg, Regensburg, Germany*
[3]*Department of Physics, Carnegie Mellon University, Pittsburgh, Pennsylvania 15213, USA*

*equal contributions

## 1. Estimation of total hydrogen coverage

In this Letter, the total hydrogen coverage on the bilayer graphene surface is estimated from fitting different regimes of the charge transport measurement result with two independent models. The different fittings yield a similar total hydrogen coverage (0.1% vs. 0.07%), which provides confidence in determining the total hydrogen concentration.

In the first method, we fit the measured gate dependence conductivity of the bilayer graphene (Fig. 2(a)) at high carrier density, where the conductivity is linear, with the following equation that describes the resonant-scattering-limited sheet conductivity [1,2]

$$\sigma_s(n) = 2 \times \frac{\pi e^2}{4h} \frac{|n|}{n_H}$$

to extract the total number of impurities and calculated the coverage based on the extracted number. Here $n$ is the carrier density (per area) in the bilayer graphene and $n_H$ is the density of atomic hydrogen. The carrier density can be extracted with $n = \frac{\varepsilon}{t}(V_g - V_{CNP})$, with $\varepsilon$ the dielectric constant of $SiO_2$ and $t = 300\ nm$ the thickness of the $SiO_2$. Due to the two additional peak features with $V_g - V_{CNP} > 0$, we only fit the curves with $V_g - V_{CNP} < 0$. Figure S1 shows the fitting for total hydrogenation of 96 s. From the fitting we obtain $n_H = 0.1\%$.

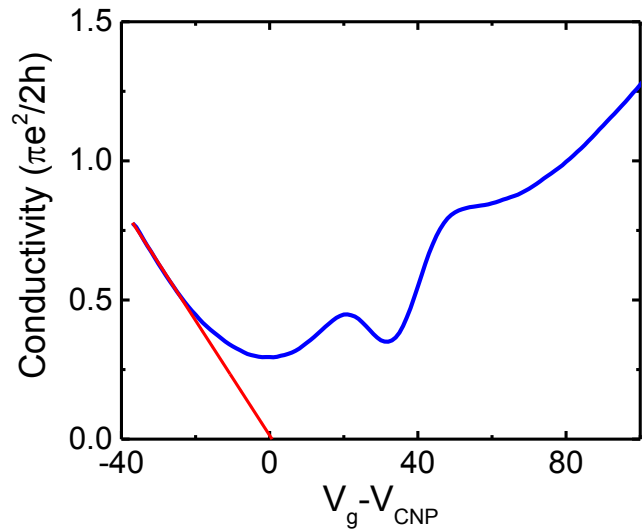

Figure S1 Gate dependent conductivity for the bilayer graphene with 96 s of total hydrogenation (blue) and the fitted linear curve (red) for extracting the total hydrogen coverage.

The second method for extracting the total hydrogen coverage is by fitting the resonant scattering peaks in the gate dependent resistance curves with the Boltzmann transport model that we developed in this letter. The result of the fitting for the 96 s hydrogenation is shown in Fig. 3(d). This method yields a total hydrogen coverage of 0.07%, which agrees well with the 0.1% coverage we determined from the first method.

2. **Evolution of α:β ratio with increasing hydrogen coverage**

One of the main features we observed in our hydrogenation experiment is that the ratio of hydrogen adatoms on different sublattices α:β changes with increasing the total hydrogen coverage. At low hydrogen coverage, most of the hydrogen adatoms concentrates on the β sublattice, while at higher coverage (~0.1%), the ratio α:β becomes close to 1:1. The change of α:β ratio was predicted by Moaied et al. [3] . They considered the adsorption of two hydrogen adatoms on bilayer graphene and explained the change in the ratio by comparing the formation energy of $E_{\beta\beta}$ and $E_{\alpha\beta}$ with different distances between the two adatoms. They also predicted that the α–β configuration is more energetically favorable when the distance is below ~10 Å (when $E_{\beta\beta} = E_{\alpha\beta}$). This corresponds to an average hydrogen coverage of ~ 3% on the bilayer graphene surface, at which point the α-β configuration becomes more energetically favorable.

The predicted 3% value is in reasonable agreement with our experimental data (~0.1%, where the ratio α:β is close to 1:1) considering that the DFT only calculated the formation energy difference of two hydrogen adatoms adsorbed under equilibrium conditions, which is a greatly simplified model compared to our experimental conditions. Other factors, such as surface diffusion of hydrogen adatoms [4,5], surface wrinkles, ripples of graphene [6,7] or inhomogeneity of charge distribution due to the SiO$_2$ substrate [8,9] can all affect the hydrogen adsorption process and the energy difference between the α–β and the β–β configuration, which are not considered in the DFT work. Furthermore, a recent comment by Bonfanti et al. [10] also shows that $\Delta_\infty$ (energy difference of single hydrogen adatom adsorption on α and β sublattice) predicted by Moaied et al. can be overestimated by as much as 50%, which implies that the Moaied et al. calculation overestimates the concentration at which the α–β configuration becomes energetically favorable (i.e. a smaller energy difference means the occupancy of the α sublattice can start at a lower concentration). These various factors provide possible reasons why the 3% estimate is higher than our experimental result. Nevertheless, the calculation by Moaied et al. is significant because it provides a conceptual framework for understanding the change of adsorption energies with hydrogen concentration as we observe experimentally.

3. **Selective removal of hydrogen adatoms from the α-sublattice**

To compare the experimental result in Fig. 4(b) with our model, we calculated the graphene resistance curves by keeping the hydrogen concentration on the β site held constant while changing the hydrogen concentration on the α site. Figure S2(a) shows the result, which is in good agreement with our experimental data.

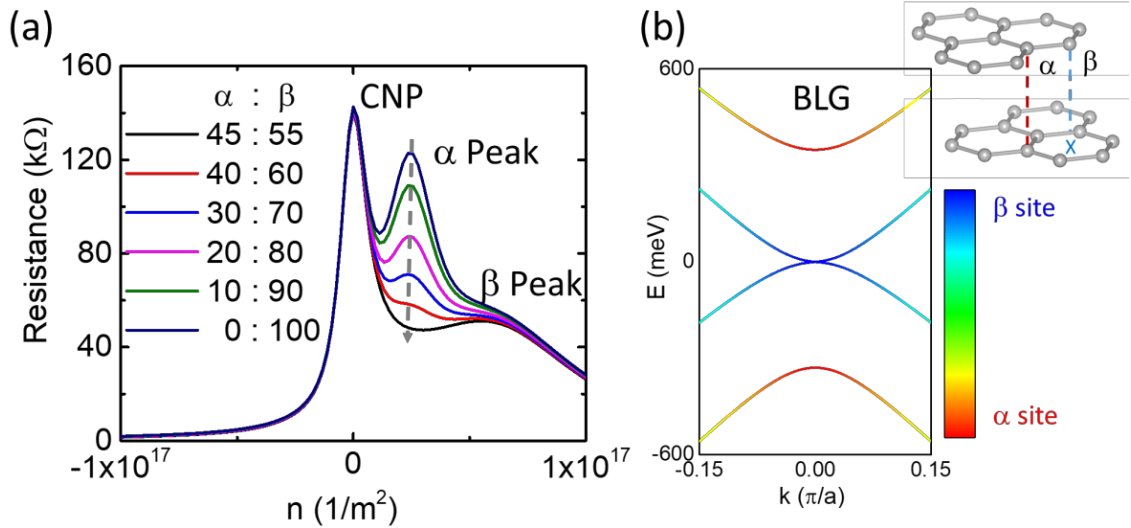

Figure S2 (a) Calculation of the bilayer graphene resistance vs. electron density (n for positive for electrons) with different α:β ratios. In the calculation, the total concentration of β site hydrogen is held constant. (b) Band structure of Bernal stacked bilayer graphene around the K point. The red (blue) color of curves shows the contribution of electronic states on the α (β) sublattice to each of the bands.

In Fig. S2(a), we also observe that the CNP resistance does not change with decreasing the total hydrogen concentration. This can be understood by considering the unique band structure of bilayer graphene [11] as shown in Fig. S2(b). Due to the Bernal stacking order in bilayer graphene, the lower energy band close to the K (K') point of graphene is formed mainly by the wave function of carbon atoms on the β sublattice, while the bands mainly formed by the α sublattice are a few hundred meV away from the CNP. When a hydrogen adatom bonds to the β sublattice, it reduces the total number of electronic states of the lower energy band, thus causing an increase in resistance at the CNP. Meanwhile if a hydrogen adatom bonds to the α sublattice, it will remove electronic states mostly from the higher energy band away from the CNP and have limited effect on the transport at the CNP. As a result, annealing and removing hydrogen from the α sublattice will also have minimal effect in the resistance of the CNP. This behavior agrees with the experimental data shown in Fig. 4(b).

In addition, while the strong reduction of the α peak in Fig. 4(b) indicates selective removal of hydrogen adsorbates from the α sublattice, we note that $V_{CNP}$ exhibits only a slight shift back toward zero upon annealing to 100 K. This is not as expected, because if a substantial fraction of the hydrogen adatom is being removed from surface, the overall charge doping of the graphene should have a significant change. The small change in $V_{CNP}$ suggests that the removal of hydrogen may proceed not only by desorption from the surface, but also from moving to the *β* sublattice or forming more complex structures, which could affect the overall charge transfer into graphene. Our transport study could not provide a definitive picture to the microscopic behavior of hydrogen adatoms during the annealing process. Future studies involving atomic scale microscopy would be essential to answer this interesting question.